\documentstyle[aps,pra,floats,twocolumn,epsfig]{revtex} 

\begin{document}
\draft 
\title{An all-optical gray lattice for atoms}
\author{H. Stecher, H. Ritsch, P. Zoller} 
\address{ Institute for
Theoretical Physics, University of Innsbruck, A-6020 Innsbruck,
Austria} 
\author{F. Sander, T. Esslinger and T. W. H{\"a}nsch}
\address{ Sektion Physik, Universit{\"a}t M{\"u}nchen, D-80799 Munich,
Germany\\and\\ Max-Planck-Institut f{\"u}r Quantenoptik, D-85748
Garching, Germany} 
\date{\today} 
\maketitle

\begin{abstract}
We create a gray optical lattice structure using a blue detuned laser
field coupling an atomic ground state of angular momentum $J$
simultaneously to two excited states with angular momenta $J$ and
$J-1$.  The atoms are cooled and trapped at locations of purely
circular polarization.  The cooling process efficiently accumulates
almost half of the atomic population in the lowest energy band which
is only weakly coupled to the light field. Very low kinetic
temperatures are obtained by adiabatically reducing the optical
potential. The dynamics of this process is analysed using a full
quantum Monte Carlo simulation.  The calculations explicitly show the
mapping of the band populations on the corresponding momentum
intervals of the free atom.  In an experiment with subrecoil momentum
resolution we measure the band populations and find excellent absolut
agreement with the theoretical calculations.
\end{abstract}

\pacs{PACS: 32.80.Pj, 33.80.Ps, 42.50.Vk }

\section{Introduction}

Remarkable progress has been made in cooling and trapping of
neutral atoms by laser fields.  An outstanding success was the
demonstration of Bose-Einstein condensation
\cite{Anderson,Hulet,Ketterle}, which was achieved by combining laser
cooling techniques with magnetic trapping and evaporative cooling. The
quest to observe quantum many body effects by manipulating atoms
merely using laser light has spurred researchers to invent new cooling
techniques.  Extremely narrow momentum distributions have been
demonstrated by velocity-selective coherent population
trapping\cite{VSCPT,Lawall,Esslinger}, Raman cooling\cite{Kasevich}
and adiabatic cooling \cite{Huletb,Philipps95,Esslingerb}. Subrecoil
cooling of atoms in mesoscopic traps has been proposed (e.g. in
Ref. \cite{Dum,darkslatt}) and recently successfully demonstrated
yielding a dramatic increase in the phase space density of the atomic
vapor \cite{Lee}.

Atoms interacting with spatially periodic light-induced potentials can
be accumulated in an array of microscopic traps. The quantum motion of
atoms in these optical lattices has been subject of extensive
research\cite{Prentiss}. The fascinating perspective that quantum
statistical effects might become observable in optical lattices has
stimulated the search for new types of optical lattices. Very
promising schemes aiming at high atomic densities are gray optical
lattices in which the trapped atoms are almost decoupled from the
light field \cite{Courtois}. This reduces the density limitation by
light induced atom-atom interactions. A one dimensional gray lattice
using a laser field in combination with a magnetic field has recently
been proposed\cite{Courtois} and theoretically investigated \cite{Petsas}. 
Experimental demonstrations of magnetic field induced dark
lattices in two and three dimensions are described in Ref.\
\cite{Hemmerich}. 

In this work we present an alternative and very simple way to
efficiently accumulate atoms in a gray optical lattice structure. Our
scheme merely uses a single frequency laser field \cite{darklatt} and
can be adapted to many alkali atoms. The adiabatic release of atoms
trapped in the optical lattice is studied both theoretically and
experimentally. We use a full quantum Monte Carlo wave function
simulation to follow the atomic evolution during the lowering of the
potential. It is shown that the band populations are indeed mapped on
the corresponding momentum intervals of the free atom, as was
suggested by Kastberg {\sl et al.} \cite{Philipps95} for adiabatic cooling
in a bright optical lattice. This population mapping is accurate if
optical pumping between different bands is negligible during the
adiabatic release. In an experiment with subrecoil momentum resolution
we demonstrate the population measurement of individual bands using
our gray optical lattice. For a wide range of parameters we find
excellent agreement between theoretically calculated band populations
and the measured values. This establishes population mapping by
adiabatic release as an experimental tool to investigate the
interaction of atoms with a spatially periodic potential.

We give a theoretical description of the gray optical lattice in
section II and calculate its eigenfunctions in section III. Numerical
simulations of the stationary energy and position distribution using
rate equations and quantum Monte Carlo wave function techniques are
presented in section IV. The adiabatic release is theoretically
investigated in section V and experimentally in section VI. Section
VII gives an outlook to experiments with high atomic densities.

\section{Optical lattice configuration}

\begin{figure}
\begin{center}
\epsfig{file=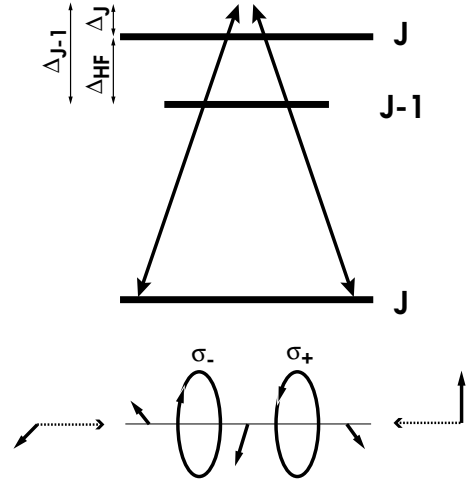,height=2.5in}
\caption{Schematic setup of the gray optical lattice.}
\label{Scheme}
\end{center}
\end{figure}

We consider a one dimensional laser field acting simultaneously on two
atomic transitions coupling an atomic ground state manifold with total
angular momentum $J_g=J$ to two different excited levels with angular
momenta $J_{e_1}=J$ and $J_{e_2}=J-1$.  The field consists of two
counterpropagating waves with mutally orthogonaly linear polarization
(lin$\perp$lin). Fig.~\ref{Scheme} illustrates the spatially varying
polarization of the standig wave. The light field shall be detuned to
the blue of both transitions and the resulting optical couplings shall have
the same order of magnitude \cite{Twolaser}.  This can be realized on
the $D_1$ line of the alkali atoms Rb, Na and Li. The two excited
levels are formed by hyperfine manifolds with the splitting
$\Delta_{HF}:=(E_{J\!e_1}-E_{J\!e_2})/\hbar$.  An essential part of
the system is the off-resonant coupling of the $J_{e_2}$ manifold with
the detuning $\Delta_{J}+\Delta_{HF}$.

The interaction of the oscillating atomic dipole with the standing
wave causes spatially periodic light shifts in the atomic ground state
manifold. In regions of purely $\sigma^+$ ($\sigma^-$) polarized light
the atoms are optically pumped into the $m$=$J_g$ ($m=-J_g$) ground
state, which is decoupled from the light field and experiences no
light shift or optical excitation. At locations of linearely polarized
light all ground state sublevels are coupled to the excited states and
are shifted towards higher energies. In this semiclassical picture we
expect that the atoms are cooled by a Sisyphus mechanism and
accumulated in dark states at locations of pure
$\sigma$-polarization. In a picture which treats the atomic motion
quantum mechanically the atomic wavefunction always has a finite
spatial extend and can not be completely decoupled from the light
field. We therefore expect the formation of a gray optical
lattice. Our situation is qualitatively similar to that of magnetic
field induced dark optical lattices, where localized dark states are
created by combining a standing wave with a magnetic field
\cite{Courtois,Hemmerich}. In the following we outline our
calculations using the example of a $J=1$ ground state.  This allows
to demonstrate the essential physics using a minimal number of Zeeman
sublevels.

A model Hamiltonian for the $J_g=1\to J_{e_1}=1,J_{e_2}=0$ system,
which includes the kinetic energy is found in Ref.\ \cite{Marte}.  To
simplify the following discussions, we assume low saturation, which
allows to adiabatically eliminate the excited atomic states
\cite{elimination}. Then the Hamiltonian can be written as:
\begin{equation}
H=\frac{P^2}{2M} + H_{AF}
\end{equation}
with
\begin{eqnarray} 
H_{AF} &=& \frac{\hbar}{2} \left( U_{1}+\frac{2}{3} U_{0}\right)\big[
\left | -1 \right \rangle \left \langle -1 \right | cos(k x+\pi/4)^2\\
\nonumber &&\qquad\qquad+\left | 1 \right \rangle \left \langle 1 \right | cos(k x-\pi/4)^2
\big]\\
\nonumber &&-\frac{\hbar}{2} \left(U_{1}-\frac{2}{3}
U_{0}\right) \big[ \left( \left | -1 \right \rangle \left \langle 1
\right |+\left | 1 \right \rangle \left \langle -1 \right | \right)\\
\nonumber &&\qquad\qquad\times cos(k x+\pi/4) cos(k x-\pi/4) \big]\\
\nonumber &&+\hbar
\frac{U_{1}}{2} \left | 0 \right \rangle \left \langle 0 \right |
\left( cos(k x+\pi/4)^2+cos(k x-\pi/4)^2 \right)
\end{eqnarray}
where $M$ is the atomic mass, $k$ the modulus of the light
wave vector and $\left | {m} \right \rangle $ the $m$-th Zeeman
substate of the ground state manifold. $U_{J\!e_j} = \Delta_{J\!e_j}
s_{J\!e_j}/2$ is the effective interaction potential (=light shift),
where $\Delta_{J\!e_j}=\omega_{\mbox{\scriptsize
laser}}-\omega_{J\!e_j}$ is the detuning and
$s_{J\!e_j}=\Omega_{J\!e_j}^2/(2\Delta^2_{J\!e_j}+\gamma_{e_j}^2/2)$
is the saturation parameter for the transitions between the ground
state manifold and the excited state manifolds. The optical pumping
rates for the ground states are proportional to the parameter
$\Gamma_{J\!e_j}=s_{J\!e_j} \gamma_{J\!e_j}$, where
$\gamma_{J\!e_j}^{-1}$ are the lifetimes of the excited levels.

\section{Optical potentials and Eigenfunctions}

\begin{figure}
\begin{center}
\epsfig{file=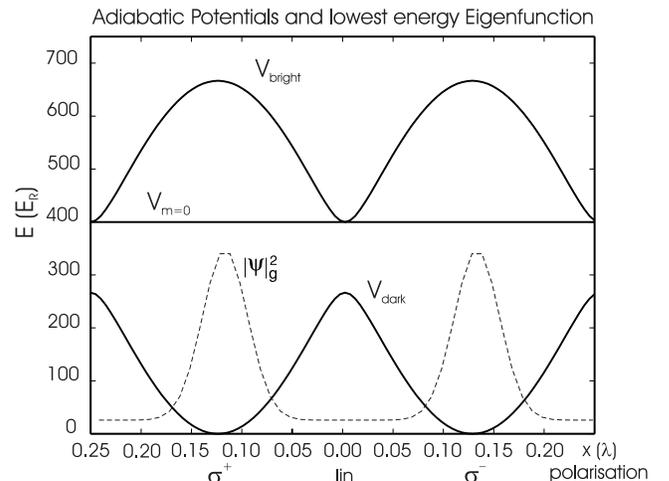,height=2.5in}
\caption{Adiabatic potentials $V_{\mbox{\tiny dark}}$, $V_{\mbox{\tiny
bright}}$ and $V_{m=0}$ (solid curves) and the atomic position
distribution of the lowest energy eigenfunction $\psi_g$ shifted by
the corresponding eigenenergy (dashed line) for the $J_g=1 \to
J_{e_j}=1,0$ system for $U_{1}=400\,E_R $ and $U_0=400\,E_R$.}
\label{AdiabaticPotentials}
\end{center}
\end{figure}

To calculate the adiabatic potentials we diagonalize the atom-field
Hamiltonian $H_{AF}$ at each spatial point $x$ separately. This yields
the spatial dependence of the eigenvalues and eigenstates.  In a
semiclassical picture these eigenvalues amount to the light shifts
experienced by an atom at rest in the corresponding states and acts as
a potential for slowly moving atoms.  Fig.~\ref{AdiabaticPotentials}
shows these adiabatic potentials for the case of a $J_g=1$ to
$J_{e_j}=1,0$ transition.  Due
to the threefold degeneracy of the unperturbed $J_g=1$ atomic ground
state, we find three adiabatic potentials. The energetically lowest
potential curve exhibits minima of zero light shift at locations of
purely circular polarization.  At the same spatial points $x$ the
maxima for the highest potential are found.  The constant potential
can be attributed to atoms in the $m=0$ state, which experience a
constant light shift in space and hence feel no (semiclassical) force.
Atoms in state $m=1$ are drawn towards regions of
$\sigma^+$-polarization, where they experience minimal (=zero) light
shift.  Atoms in the $m=-1$ are repelled from this $\sigma^+$-area
(uppermost curve).  The magnitude of both forces is comparable and an
atom moving in the $\sigma^+$-region spends most of its time in the
weakly coupled $m=1$ state. Therefore the semiclassical picture
predicts a trapping force towards areas of circular polarization. In
addition the corresponding optical pumping rates show the right
spatial dependence to provide a Sisyphus cooling mechanism for moving
atoms, as they are pumped into the locally less coupled (i.e.  the
energetically lower lying) states.

We now include the motional degrees of freedom and calculate the
eigenstates of the full Hamiltonian $H$. The diagonalization is
performed numerically on a discrete spatial grid extending over
several wells of the optical potential with periodic boundary
conditions. The calculated discrete eigenstates correspond to the
energy bands of the full spatially periodic lattice structure.  At
energies above the well depth we find delocalized wavefunctions
(unbound states) whereas at low energies the states are well localized
(bound states).  The atomic position distribution $\left |
\psi_g(x)\right |^2$ of the energetically lowest eigenstate is plotted
with a dashed line in Fig.~\ref{AdiabaticPotentials}.  The chosen
vertical offset is the ground state energy.  As expected we find a
strong localization of the wavefunction at the potential minimum of
each well.  Its width is a fraction of the optical wavelength and
depends on the detuning and the intensity of the light field.  The
momentum spread associated with the finite size of the optical wells
prevents the existence of eigenstates exactly decoupled from the light
field. Nevertheless we find, that the localized states of lowest
energy exhibit only a very small optical coupling to the two excited
levels and hence they have correspondingly low light shifts and
energies.

The energy intervals between the lowest states are considerably
larger than the recoil energy and reach up to a few hundred recoil
shifts. Experimentally they determine the Raman resonances found in
fluorescence or in probe absorption spectra \cite{Prentiss}. With
increasing energy the interval sizes decrease due to the
anharmonicity of the potential. This leads to inhomogeneous broadening
of the probe absorption resonances.

The tunnel coupling between two equivalent states of neighbouring
potential wells can be estimated from the eigenenergies of a two well
calculation. The coupling leads to an energy splitting between the
corresponding symmetrical and antisymmetrical states. For the
parameter regions and time scales we consider in the following, the
tunnel coupling between the bound low energy states of the single
wells is so small, that we can view the potential wells as
independent. For situations with finite tunnel coupling it would be
advantageous for calculations to use Bloch eigenfunctions of the
optical lattice as the numerical basis \cite{Marte}.

\section{Energy and position distributions}

We now calculate the steady state distributions of atoms in the
optical lattice. They allow to judge the effectiveness of the involved
cooling and confinement mechanisms on the basis of quantities as the
mean energy, the position spread or the population of the
energetically lowest bound states.

The inclusion of the spontaneous scattering of photons in addition to
the coherent atom-laser field dynamics implies a dynamical
redistribution of the population among the various atomic states
eventually leading towards a steady state\cite{rate,Marte,Dum}.  We
calculate the steady state population distribution in a similiar
approach as demonstrated in Refs.\ \cite{Courtois,darkslatt,Marte}
using two methods: a rate equation apporach based on the Raman
transition matrix elments between the eigenstates
\cite{rate,darkslatt} and a Quantum Monte Carlo wave function
simulation technique (QMCWFS) relying on a Bloch-state expansion of
the atomic wavefunctions\cite{Marte,Dum}.  This allows to
realistically treat periodic spatial geometries still maintaining a
numerically tractable grid size. We expect the rate equations to give
fairly accurate results for the bound states, where the tunnel
coupling is small, and less accuracy for states with energies near and
above the potential well depth.

\begin{figure}
\begin{center}
\epsfig{file=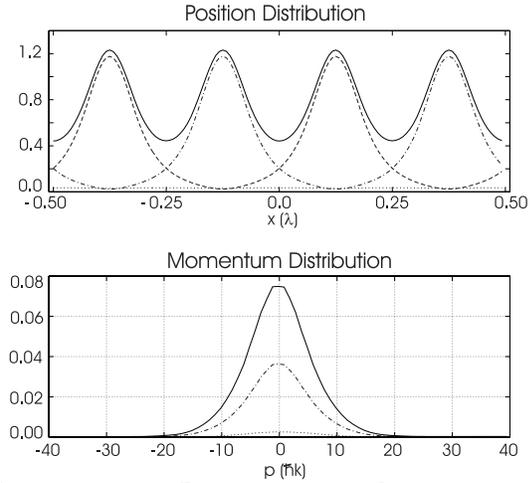,height=2.5in}
\caption{Position and momentum distribution for $U_{1} =
200\,E_R$, $U_{0} = 100\,E_R$, $\Gamma_{1}= 3\,E_R/\hbar$ and $\Gamma_{0}=
1\,E_R/\hbar$ for the $J_g=1 \to J_{e_j}=1,0$ system.
The uppermost lines give the total distributions, while the
lower lines give the partial contributions of the various Zeeman
substates. The dotted curves correspond to the $m=0$ substate. The 
momentum distribution is normalized and given in units
of $(\hbar k)^{-1}$, the position distribution is periodic and given in
arbitrary units.}
\label{Distributions}
\end{center}
\end{figure}

The stationary position and momentum distributions obtained by the
QMCWFS are shown in Fig.~{\ref{Distributions}}. The upper plot shows
the total position distribution (solid curve) as well as the
contributions of the various Zeeman sublevels (dashed curves). As
expected the $m=0$ state is almost not populated and shows no
spatial variation. The localization towards points of
purely circular polarization is state selective. We find better
localization with increasing field strength.  Due to the strong
spatial confinement to roughly $\Delta x =\lambda/8$, we find a
correspondingly large width of the momentum distribution of $\delta p
> 8 \hbar k$, which is shown in the lower plot of Fig.\
{\ref{Distributions}}.  The obtained results are in excellent
agreement with the results of the rate equation model as shown below.

The optical pumping rates scale with the parameters $\Gamma_{J\!e_j}$,
which depend on the lifetimes of the excited states and on the
detuning and the intensity of the laser field.  Their relative
magnitude can be tailored to a large extend by a suitable choice of
these two laser parameters. The possible ranges of obtainable light
shifts $\propto U_{J\!e_j}$ and optical pumping rates
$\propto\Gamma_{J\!e_j}$ may be limited by the available laser
intensity and by the magnitude of the atomic hyperfine splitting. An
unfavourable choice of parameters can lead to additional unwanted
off-resonant couplings to other atomic hyperfine levels.

\begin{figure}
\begin{center}
\epsfig{file=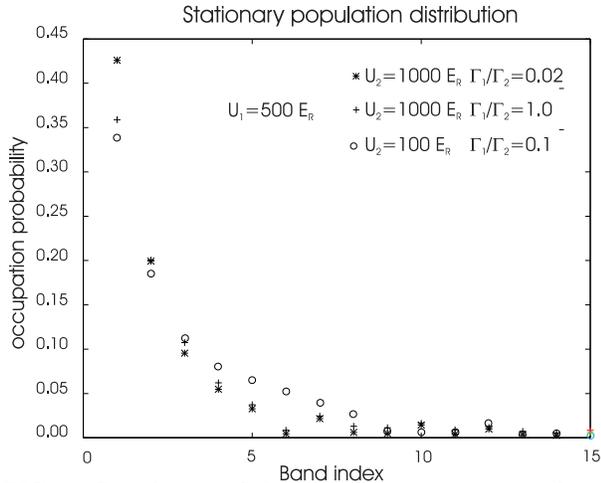,height=2.5in}
\caption {Population of the low energy eigenstates for the $J_g=2 \to
J_{e_j}=2,1$ system for
various lightshifts $U_{2}$ and ratios of the optical pumping rates
$\Gamma_1/\Gamma_2$}
\label{Populations}
\end{center}
\end{figure}

The influence of the lightshift $U_{J\!e_j}$ and optical pumping rate
parameters $\Gamma_{J\!e_j}$ is demonstrated in Fig.\
\ref{Populations} for the $J_g=2\to J_{e_j}=2,1$ transition of the
$^{87}$Rb $D_1$-line. We chose this transition as an example, because
the hyperfine splitting allows a wide range of this parameters.  The
figure shows the occupation probabilities for three sets of
parameters.  The data points marked with '*' correspond to a
situation, where the coupling to the excited state manifold
$J_{e_1}=2$ is near resonant and the coupling to the $J_{e_2}=1$
manifold is off resonant. The data points marked with '+'
correspond to a situation with couplings of equal strengths. The
highest ground state occupation probability of 43\% was found in the
first case with $U_1=500\,E_R$, $U_2=1000\,E_R$ and
$\Gamma_1/\Gamma_2=0.02$. In all cases a large fraction of the atomic
population is concentrated in the lowest few (gray) states. This
corresponds to strong spatial localization of the atoms, a result,
which we obtain for the $J_g=1\to J_{e_j}=1,0$ transition also by the
QMCWFS. The computation time required for QMCWFS for transitions with
$J_g>1$ is so high, that we selectively performed Monte Carlo simulations
to verify the rate equation results. For transitions
with higher angular momentum $J_g$ the population of the ground state
(lowest band) is higher, because the relative magnitude of the
Clebsch-Gordan coefficients for the states with $m=\pm J_g$ and
$m=\pm(J_g-1)$ changes.

The energy distributions in the example strongly deviate from thermal
distributions of the same mean energy. Therefore we do not have a
thermal equilibrium state to which one could consistently attribute a
temperature. The ground state population for the case with
$U_2=1000\,E_R$ and $\Gamma_1/\Gamma_2=0.02$ is $\approx\!20\%$ higher
than for a thermal distribution of same mean energy $\langle E
\rangle\approx130\,E_R$.  So the temperature deduced from the relative
populations of the lowest two levels k$_{\rm B} T_{12}= 76\,E_R$
differs by $\approx 25\%$ from the corresponding value k$_{\rm B}
T_{34}= 96\,E_R$ obtained from levels $n=3$ and $n=4$. The
disagreement becomes more significant for higher $n$.
 
\begin{figure}
\begin{center}
\epsfig{file=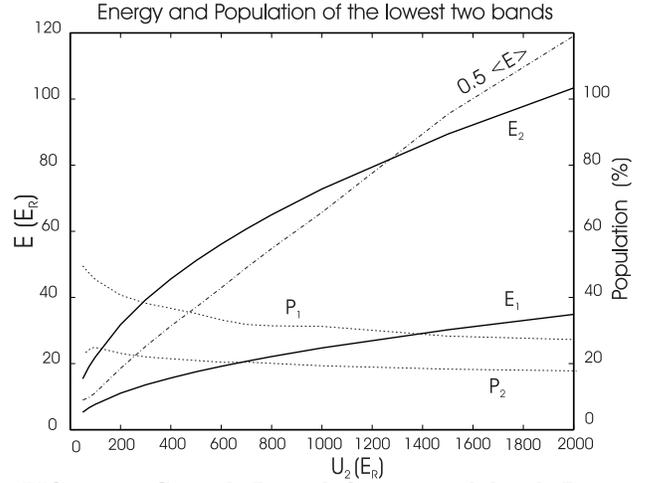,height=2.5in}
\caption { Ground $P_1$ and first excited band $P_2$ populations
 (short dashed curves), the corresponding energies $E_1$, $E_2$ and
 the mean energy $\langle E \rangle$ (dot dashed line) for the $J_g=2
 \to J_{e_j}=2,1$ system as a function of the laser intensity $\propto
 U_{2}$ for fixed ratios of the lightshift potentials
 $U_{1}/U_{2}=1/3$ and of the optical pumping rates
 $\Gamma_1/\Gamma_2=0.1$.}
\label{Popground}
\end{center}
\end{figure}

The mean energy $\langle E \rangle$ of the atoms, which is e.g.\ of
central importance for loading of a purely magnetic trap, and the
stationary value of the atomic population $P_1$ of the lowest energy
band are key quantities with respect to a possible direct observation of
quantum statistical effects. The dependence of $\langle E \rangle$,
$P_1$ and $P_2$ on the light shift $U_{2}$ for fixed ratios
$\Gamma_1/\Gamma_2=0.1$ and $U_1=U_2/3$ is shown in Fig.\
{\ref{Popground}}.  This case corresponds to a laser tuned far to the
blue of the $J_g=2\to J_{e_j}=2,1$ transitions.  A probability of
$P_1=45\%$ to find the atom in the lowest energy eigenstate is achieved for
parameters well in the reach of experimental capabilities.

\section{Adiabatic Release}\label{release}

It has been suggested in recent experimental
work\cite{Philipps95,Esslingerb} that the populations of an optical
lattice can be directly experimentally measured by an adiabatic
release of the atoms from the lattice and a subsequent measurement of
the resulting atomic velocity distribution. If the release is fully
adibatic, the atoms from the lowest (first) band will be mapped
exactly to the momentum interval between $-\hbar k$ and $+\hbar k$
and the second band will be mapped to the intervals $-2\hbar k$ to
$-\hbar k$ and $\hbar k$ to $2\hbar k$. The $n$-th band will be mapped
on the $-n\hbar k$ to $-(n-1)\hbar k$ and the
$(n-1)\hbar k$ to $n\hbar k$ intervals. Nonadiabaticity and incoherent
redistribution of the atoms during the release will alter the mapping
and the assignment of the band populations of the lattice and the
momentum intervals of the free atoms will be less accurate. 
We have performed a QMCWFS with a time dependent
laser intensity to quantitatively verify the mapping between the stationary
population distribution of the lattice and the momentum distribution
of the free atom.  All the effects of nonadiabaticity and incoherent
spontaneous redistribution of the atoms during the release are fully
accounted for in this model.

\begin{figure}
\begin{center}
\epsfig{file=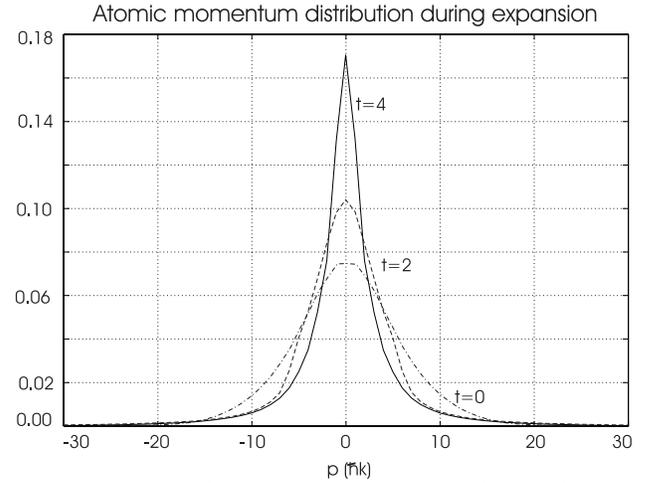,height=2.5in}
\caption { Normalized atomic momentum distribution during various
stages of the field turnoff $t=0$, $t= 2\,\tau_R $ and $t= 4\,\tau_R$ for
a $J_g=1 \to J_{e_j}=1,0$ system and for the same parameters as in
Fig.~\protect\ref{Distributions}.  The optical potential decreases as
$U_{J\!e_j}(t) = U_{J\!e_j}(0) exp(-{1 \over 2} (t/\tau_R)^2)$. The
units on the vertical axis are $(\hbar k)^{-1}$.}
\label{Adiabatic}
\end{center}
\end{figure}

The time evolution of the atomic momentum distribution is shown in
Fig.~\ref{Adiabatic} for an gradual turnoff of the lattice field.  For
$t=0$ we start with the steady state distribution, which we calculated
by the rate equation approach disussed above. The atoms are localized
in the lattice and correspondingly the momentum distribution is
broad. When the lattice field intensity is reduced, the momentum
distribution becomes narrower due to adiabatic cooling. For the given
example the optical potential varies as $U_{J\!e_j}(t) = U_{J\!e_j}(0)
exp(- (t/\tau)^2)$ as it would occur if an atom leaves a Gaussian beam
transversly at a constant velocity. The chosen time scale of the
turnoff $\tau = \sqrt{2}\,\tau_R$ (where $\tau_R=2\pi\hbar/E_R$ is the
recoil time) is long compared to the oscillation period of atoms in
the lattice wells but short compared to the lifetime of the lowest
dark levels. The lightshift parameters $U_1=200\,E_R$, $U_0=100\,E_R$
and the optical pumping rates $\Gamma_1=3\,E_R/\hbar$ and
$\Gamma_0=1\,E_R/\hbar$ are the same as in Fig.~\ref{Distributions}.

The narrowing of the momentum distribution continues for $t\gg\tau$
down to a width of roughly one recoil momentum.  From the resulting
momentum distribution we calculate the probability $W_N$ to find an
atom in the interval between $-N \hbar k$ and $+N\hbar k$. We find
$W_1= 37,9\%$, $W_2=56,4\% $ and $W_3=68,1\% $. This agrees to within
less than $1\%$ with the corresponding initial (steady state)
populations of the lattice $P_1=37,4\%$, $P_1+P_2=56,3\%$ and 
$P_1+P_2+P_3=67,9\%$.
A comparison for higher momenta is limited by the accurracy of the
rate equation approach.

\section{Experiment}

To demonstrate the experimental feasibility of the proposed method to
measure the band populations of the optical lattice, we performed an
experiment with the apparatus described in
Refs.\cite{Esslinger,Esslingerb}. Our gray lattice scheme is realized on the
$J_g=3 \to J_{e_j} = 3,2$ transitions of the $^{85}$Rb $D_1$-line
(where $J$ is the total angular momentum of the atom).

A pulsed beam of cold Rubidium atoms is directed vertically downwards
and crosses a standing wave field (lattice field), which induces the
optical potentials of the dark lattice.  The atoms are cooled into the
lattice sites and are then gradually released from the optical
potential due to the Gaussian shape of the lattice field.  Below the
lattice field the momentum distribution of the atoms is measured with
a resolution of one third of the photon recoil.

The lattice field is induced by a standing wave oriented along the
$x$-axis and has a frequency tuned $\Delta_3=26\,\gamma_3$ to the blue
of the $J_g=3 \to J_{e_1}=3$ transition. The hyperfine-splitting
between the two excited states of the $D_1$-line is $\Delta_{HF} =
65\,\gamma_3$ and the detuning of the second excited states is
$\Delta_{2} = 91\,\gamma_3$.  The incoming beam of the standing wave
is linearly polarized along the $z$-axis and the reflected beam is
polarized along the y-axis.  The Gaussian waists of the beams are
$w_z=1.35$\,mm in $z$-direction and $w_y=0.48$\,mm in $y$-direction.
This corresponds to a 0.4\,ms time of flight of the atoms (3.2\,m/s)
through the waist $w_z$.  The region of the lattice field is shielded
against stray magnetic fields to well below 0.5\,mG.  A second
standing wave overlaps the lattice field.  It is tuned to the $J=2 \to
J=3$ transition of the $^{85}$Rb D$_2$-line and optically pumps the
atoms into the $J=3$ groundstate.

To determine the atomic momentum distribution a pinhole, $75\,\mu$m in
diameter, is placed 5\,mm below the standing wave axis and the spatial
distribution of those atoms passing through the pinhole is measured
9.6\,cm further down by imaging the fluorescence in a sheet of light.
The sheet of light is formed by a standing wave which is resonant with
the $J=3 \to J=4$ closed cycle transition of the D$_2$-line.  For each
set of parameters we accumulate 200 single shot images and subtract
the separately measured background.  To obtain a one dimensional momentum
distribution in $x$-direction we integrate the two dimensional
distributions along the $y$-axis.

\begin{figure}
\begin{center}
\epsfig{file=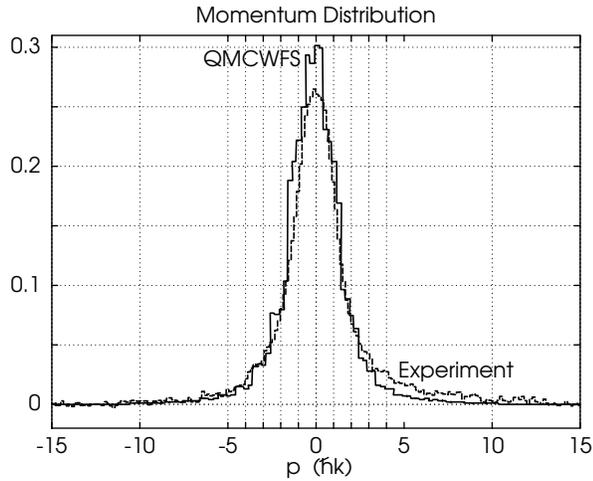,height=2.5in}
\caption { Normalized momentum distribution of the atoms after their
adiabatic release from the lattice. The QMCWF simulation corresponds
to the solid line, while the dashed line gives the corresponding
experimental result for $\Delta_{3} = 26\,\gamma_3$ and $U_{3}=
\gamma_3/4$. The units on the vertical axis are $(\hbar k)^{-1}$.}
\label{mom}
\end{center}
\end{figure}

Fig.~\ref{mom} shows an experimentally measured momentum distribution
(dotted curve) in comparison with the corresponding distribution
obtained from the QMCWFS (solid curve). The initial distribution for
the Monte Carlo simulation was the steady state distribution of the
optical lattice in the center of the Gaussian beam. The calculation
was performed as explained in section \ref{release} but using a finer
momentum grid.  The plotted curves are in good agreement except for a
small fraction of atoms. In the experiment more atoms are found at
higher velocities and less at low velocities. This can be related to
the finite interaction time. The atoms in the experiment do not reach
the steady state distribution in the center of the Gaussian beam and
some fast atoms are not yet cooled into the lattice wells.

\begin{figure}
\begin{center}
\epsfig{file=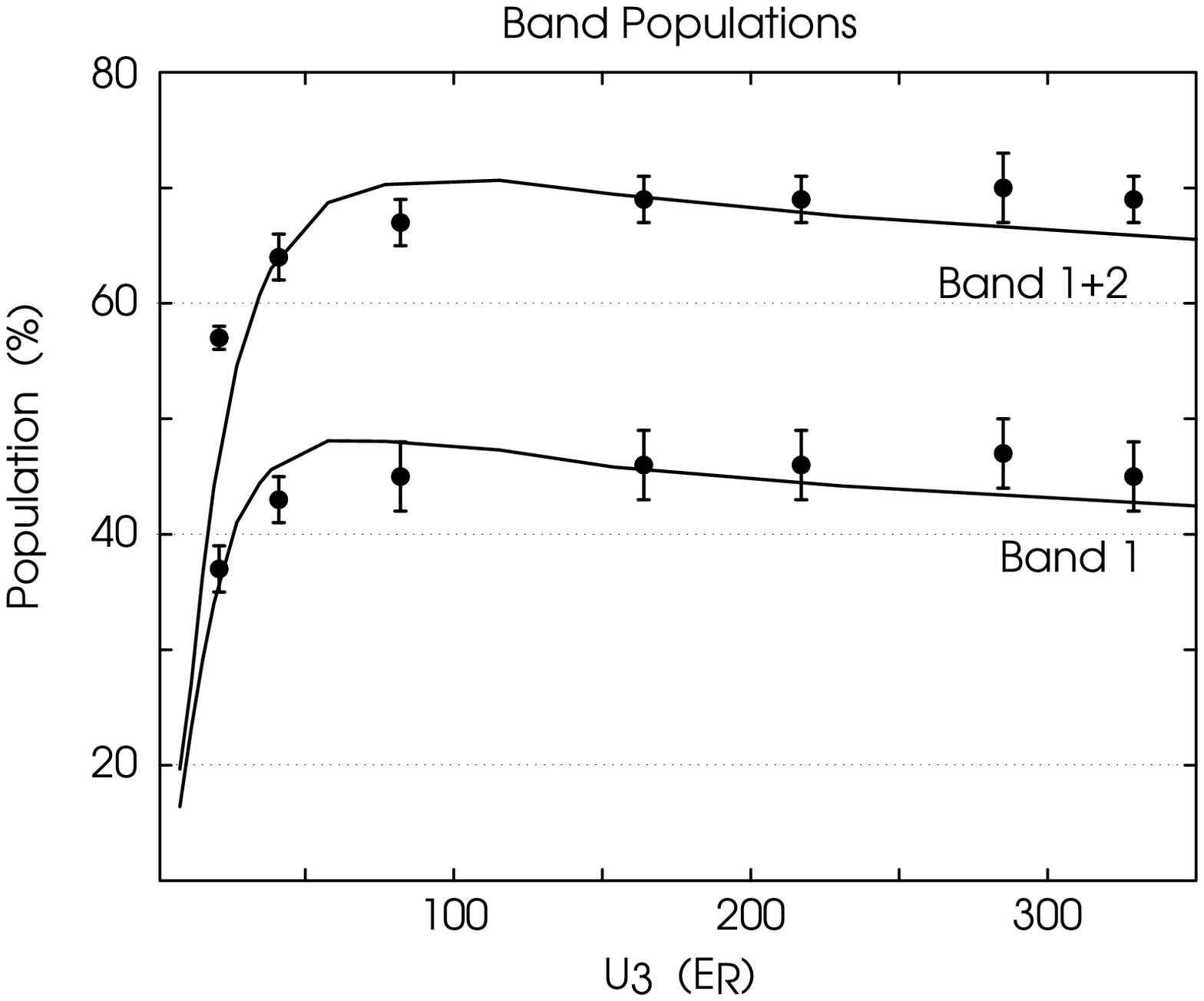,height=2.5in}
\caption { Population of the lowest (lower curve) and lowest two
(upper curve) bands of the optical lattice as a function of the
lightshift as obtained by quantum rate equations (solid lines) in
comparison to the experimentally measured results (data points) for
the $J_g=3 \to J_{e_j}=3,2$ transition of the $^{85}$Rb
$D_1$-line. The detuning was chosen to be $\Delta_3=26\,\gamma_3$ to the blue
of the $J_g=3 \to J_{e_1}=3$ transition.}
\label{popula}
\end{center}
\end{figure}

For the experimental data we count the number of atoms detected in the
momentum intervals $-\hbar k$ to $+\hbar k$ and $-2\hbar k$ to
$+2\hbar k$ corresponding to the populations in the lowest energy band
and the two lowest energy bands, respectively.  These experimentally
obtained populations are plotted in Fig.~\ref{popula} with data
points versus the lightshift $U_3$ on the $J_g=3 \to J_{e_1}=3$
transition in the center of the Gaussian beam.  The data points were
recorded for several intensities and fixed detuning.  The solid line
represents the (steady state) band populations in the lattice
calculated for the center-of-beam parameters using the rate equation
approach.  The experimentally measured populations and the calculated
steady state populations agree within 5\% over the full investigated
range of parameters. This is remarkable, because the calculation was
based only on the detunings and the intensities in the center of the
Gaussian beam and the comparison involves no fit parameter. The small
deviations for high lightshift parameters $U_3$ towards higher ground
state population for the experimental values can be attributed to a
small but finite spontaneous emission probability during the release
of the atoms, which especially effects the energetically higher lying
less dark states and which transfers additional population to the
ground state.

\section{Outlook}

An interesting field for future experimental and theoretical work is
the interaction of a high density atomic sample \cite{Ellinger} with a
periodic optical potential \cite{Goldstein}.  It has been suggested
\cite{Wilkens,Holland} that quantum statistical effects may tend to
cluster bosonic atoms within a single well of an optical potential and
that laser-like sources for matter waves might become feasible. In our
theoretical work we have not included any atom-atom interaction. Thus
we can only speculate that the low photon scattering rate for atoms
trapped in a gray optical lattice leads to much higher achievable
atomic densities than for the case of bright optical lattices. This
optimism is based on the assumption that atom-atom perturbations by
dipole-dipole interaction and reabsorption of scattered photons are
much reduced for atoms trapped in a dark state.  The atomic densities
necessary for the observation of quantum statistical effects are
considerably lower in steep potentials compared to the cases of wide
traps or free particles. The optical potentials
of gray optical lattices can have an energy spacing much larger than
the recoil energy and the trapped atoms can be cooled to mean
energies of the same order of magnitude. We reached in one dimension a
ground state occupation probability of $\approx\!50\%$, so that ---
extended to three dimensions --- already two atoms located in the same
potential well are sufficient for quantum statistical effects to become
relevant. This might be observable even at average filling factors below one
atom per lattice site.

In conclusion, we have theoretically and experimentally studied a gray
optical lattice structure which combines a low photon scattering rate
with a high population in the lowest energy band. The lattice is
formed by coupling an atomic ground state to two excited states. The
atoms are trapped at locations of purely circular polarisation which
allows an extension of the scheme to two and three dimensions using
the same field configurations as for bright optical lattices. We have
numerically simulated the dynamics of atoms adiabatically released
from the optical potential and the mapping of the band populations on
the corresponding momentum intervals. The quantitative agreement with
the band populations measured in the experiment shows that adiabatic
release is a promising tool to study the density dependence of the band 
populations in an optical lattice.

\section{Acknowledgements}

We wish to thank P.~Marte, A.~Hemmerich, T.~Pellizzari, S.~Marksteiner
and K.~Ellinger for many helpful and stimulating dicussions.  This
work was supported by the {\"O}sterreichischer Fonds zur F{\"o}rderung
der wissenschaftlichen Forschung under grants No. S6506/S6507 and by
the Deutsche Forschungsgemeinschaft.

% Figure captions

\begin{references}
\bibitem {darklatt} The optical lattice described in Ref.\
\cite{Esslingerb} is induced by coupling a ground state with angular
momentum J to two excited states with angular momenta $J$ and $J+1$. For
this case the localized atoms are off-resonantly coupled to the
excited state with angular momentum $J+1$ and decoupled from the excited
state with angular momentum $J$.
\bibitem{Twolaser} It is also possible to build up the lattice
structure with two laser fields, each coupling one of the two excited
states to the ground state. The frequency difference between this two
fields should be small, so that their relative phase mismatch can be
ignored over the size of the experimental setup. Otherwise one obtains
an additional superlattice structure \cite{darkslatt,Grimm}.
\bibitem{elimination} This assumption is not essential here and might
not be the experimentally optimal case, but it greatly reduces the
calculational effort.
\end{references}
\end{document}